\documentclass{caosp302}

\usepackage{txfonts}\usepackage{balance}
\usepackage{changebar,graphicx}

\def\teff{$T\rm_{eff }$}

\articleNo{123}
\pubyear{2005}
\volume{35}
\volnumber{3}
\firstpage{1}
\received{December 1, 2007}
\accepted{, 2007}
\begin{document}

%
\hauthor{M.\,Vick and G.\,Michaud}

\title{The Effects of Mass Loss on the Evolution of Chemical Abundances in Fm Stars}


%
\author{
        M.\,Vick \inst{1,} \inst{2} 
      \and 
        G.\,Michaud \inst{1}   
       }

%
\institute{
D\'epartement de physique --
Universit\'e de Montr\'eal, Montr\'eal, Qu\'ebec, Canada,
H3C 3J7
         \and 
Universit\'e Montpellier II, GRAAL-UMR5024/IPM (CNRS),
Place 
Eug\`ene-Bataillon, 34095 Montpellier, France
\newline\email{mathieu.vick@umontreal.ca, michaudg@astro.umontreal.ca}
	          }

\date{December 1, 2007}

\maketitle

\begin{abstract}
Self-consistent stellar models including all effects of atomic diffusion and radiative accelerations as well as mass loss
are evolved from the pre main sequence for stars of 1.35-1.5\,M$_{\odot}$ at solar metallicity (Z=0.02). A mass loss rate similar to the solar mass loss rate is sufficient to reproduce observations of the star $\tau$UMa. We discuss the effect of mass loss on the iron convection zone that naturally appears beneath the main hydrogen convection zone of these stars. We also find that the effects of mass loss can be distinguished from those caused by   turbulence, but are nevertheless able to explain the particularities of the AmFm phenomenon.
\end{abstract}

%
\section{Introduction}
\label{intr}

Since 1970 it is generally agreed that atomic diffusion driven by radiative accelerations plays a role in creating the anomalous surface abundances of F, A and B stars (Michaud 1970). However, some 40 years later, many questions remain as to the exact behavior of many physical processes within the stable envelopes of these stars. In fact, for Fm stars, two competing scenarios which have each had their share of success are presently being confronted to observations. The  ``classical'' scenario Watson (1971) suggests that separation occurs below the H-He convection zone. In this framework, models which only consider atomic diffusion without extra mixing generate predicted anomalies that are 3-5 times larger than the ones observed (Turcotte et al. 1998), thus implying that there is at least one competing process that slows the effects of separation. This lead to more recent models (Richer et al. 2000, Richard et al. 2001) in which turbulence enforces mixing down to about 200\,000\,K. In these models, it is implied that separation occurs deeper in the star. 

Like turbulence or rotation, mass loss is another macroscopic process that can reduce inhomogeneities on the surface of these stars. However, until now, only static stellar models have included the effects of mass loss (Michaud et al. 1983, Alecian 1996 for Ca and LeBlanc et Alecian 2007 for Sc) with the latter paper showing that the actual depth at which separation occurs is still uncertain.  

With self-consistent models of Fm stars (6500\,K\,$\leq$\,\teff\,$\leq$\,7000\,K) we will show that mass loss can reduce predicted abundances to the observed levels. The first aim is to see to what extent observations can constrain the importance of mass loss and if its effects can be deciphered from the ones encountered with turbulence. We will also discuss the implications of our models on the depth of chemical separation.    

\section{Evolutionary Models}

The following is a continuation of the Montreal stellar model development project (Richard et al., 2001 and references therein). The evolutionary calculations take into detailed account the time-dependent variations of 28 chemical species and include all effects of atomic diffusion and radiative accelerations. These are the first fully self-consistent stellar models which include mass loss. Models were calculated for 1.35\,M$_{\odot}$, 1.40\,M$_{\odot}$,1.45\,M$_{\odot}$ and 1.50\,M$_{\odot}$. All models have evolved from the homogeneous pre-main sequence phase with a solar metallicity (Z=0.02). The mass loss rates considered varied from 1\,$\times$\,10$^{-14}$M$_{\odot}$yr$^{-1}$ to 3\,$\times$\,10$^{-13}$M$_{\odot}$yr$^{-1}$. The mass loss is considered spherical, chemically homogeneous and weak enough not to disturb burning in the core or stellar structure. The net effect is the appearance of an outward flowing wind which is represented as an advection term in the transport equation. Due to numerical instabilities resulting from very large advection velocities in the surface convection zone, some adjustments must be made in order to avoid convergence problems. The method is well described in Charbonneau (1993). The transport equation then becomes:

%
%
\begin{eqnarray}
\nonumber\rho\frac{\partial c}{\partial t}&=&-\nabla\cdot\mbox{[}-\rho D{\bf\nabla}\ln c+\rho({\bf U}+{\bf U}_{w})c\mbox{]}\\
&&+\rho (S_{nuc}+S_{w})c,
\label{charb2}
\end{eqnarray}
with a Neumann condition (no flux) imposed at the surface and with ${\bf U}_{w}$ and $S_{w}$ defined as:    
\begin{equation}
{\bf U}_{w}=\left\{\begin{array}{l}w_{w}{\bf\hat e_{r}}$ $ \mbox{under the SCZ,}\\ 0$ $ \mbox{in the SCZ;}\end{array}\right. 
\label{uw}
\end{equation}
\begin{equation}
S_{w}=\left\{\begin{array}{l}0$ $ \mbox{under the SCZ,}\\ \frac{\dot M}{M_{ZC}}$ $ \mbox{in the SCZ.}\end{array}\right.
\label{sw}
\end{equation}
Here, $c$ is the time and depth dependent concentration, $\rho$ is density, $D$ is the total diffusion coefficient, $U$ is the total velocity field, $U_w$ 
is wind velocity, $M_{CZ}$ is the mass of the SCZ, $\dot M$ is the mass loss rate, $S_{nuc}$ is a source/destruction term linked to nuclear reactions and $S_w$ is a destruction term linked to mass loss.  

\section{The Effects of Mass Loss on Surface Abundances}
One of the effects of mass loss is to drag elements which tend to sink. Because the diffusion velocity must be greater than the wind velocity for separation to occur, the greater the mass loss the deeper we have to look to see any effects of separation. On the other hand, elements which are naturally supported by the radiation field will be pushed into the surface convection zone and evacuated through the stellar surface. As seen in Figure 1, models in which only atomic diffusion is included lead to larger surface abundances anomalies than in the presence of mass loss. For instance,  after 0.8 Gyr of evolution predicted anomalies (in terms of the original abundance) for the plotted elements are respectively, without  mass loss and with a mass loss rate of 5\,$\times$\,10$^{-14}$M$_{\odot}$yr$^{-1}$, $\times$0.2 and $\times$0.3 for Li, $\times$0.25 and $\times$0.3 for Ca, $\times$4 and $\times$1.5 for Fe. We also see that for the given stellar mass, a mass loss rate of 5\,$\times$\,10$^{-14}$M$_{\odot}$yr$^{-1}$ is sufficient to reduce anomalies by a factor of 1.5 to 3, and a mass loss rate of 3\,$\times$\,10$^{-13}$M$_{\odot}$yr$^{-1}$ practically flattens the surface abundances. 
\begin{center}
\begin{figure*}[t!]
\includegraphics[scale=.63]{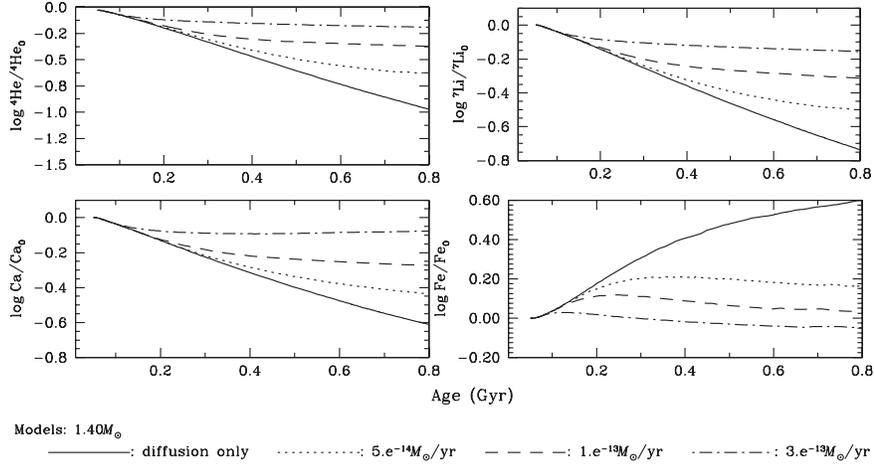}
\caption{\footnotesize Evolution of surface abundances ($^4$He, Li, Ca and Fe) for  1.4\,M$_{\odot}$ models with different mass loss rates as well as with atomic diffusion only. It is clear that a slight change in the mass loss rate can have an important effect on abundance anomalies.
}
\end{figure*}
\end{center}

\subsection{The 1.5\,M$_{\odot}$ models}

\begin{figure*}[t!]
\begin{center}
\includegraphics[scale=.61]{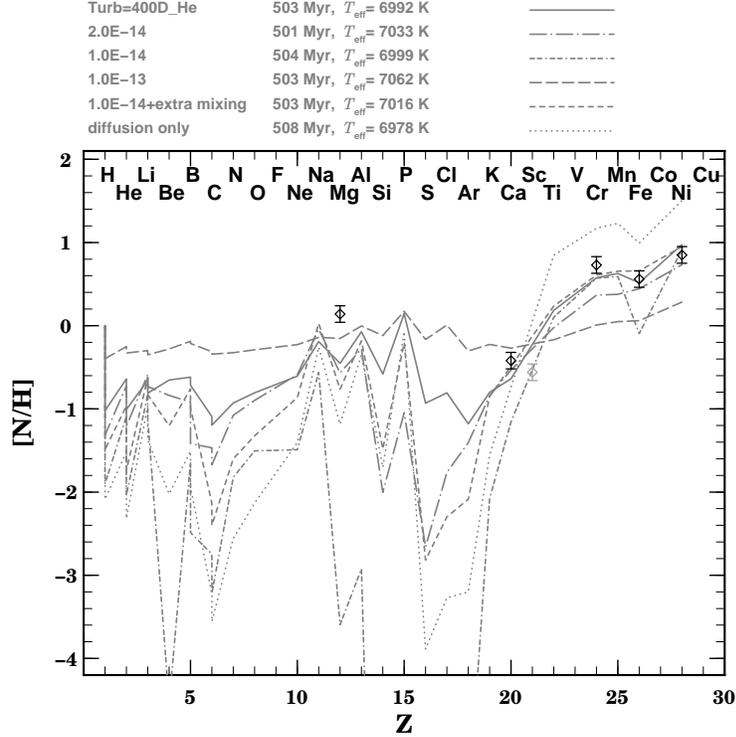}
\caption{\footnotesize
Observed surface abundances of $\tau$UMa (diamonds) compared to 1.5\,M$_{\odot}$ models at 500 Myr. Scandium is not include in our calculations. Models with mass loss are designated by their respective rate (e.g 1.0E-14$\rightarrow$1\,$\times$\,10$^{-14}$M$_{\odot}$yr$^{-1}$).}
\end{center}
\end{figure*}
\begin{figure*}[t!]
\begin{center}
\includegraphics[scale=.30]{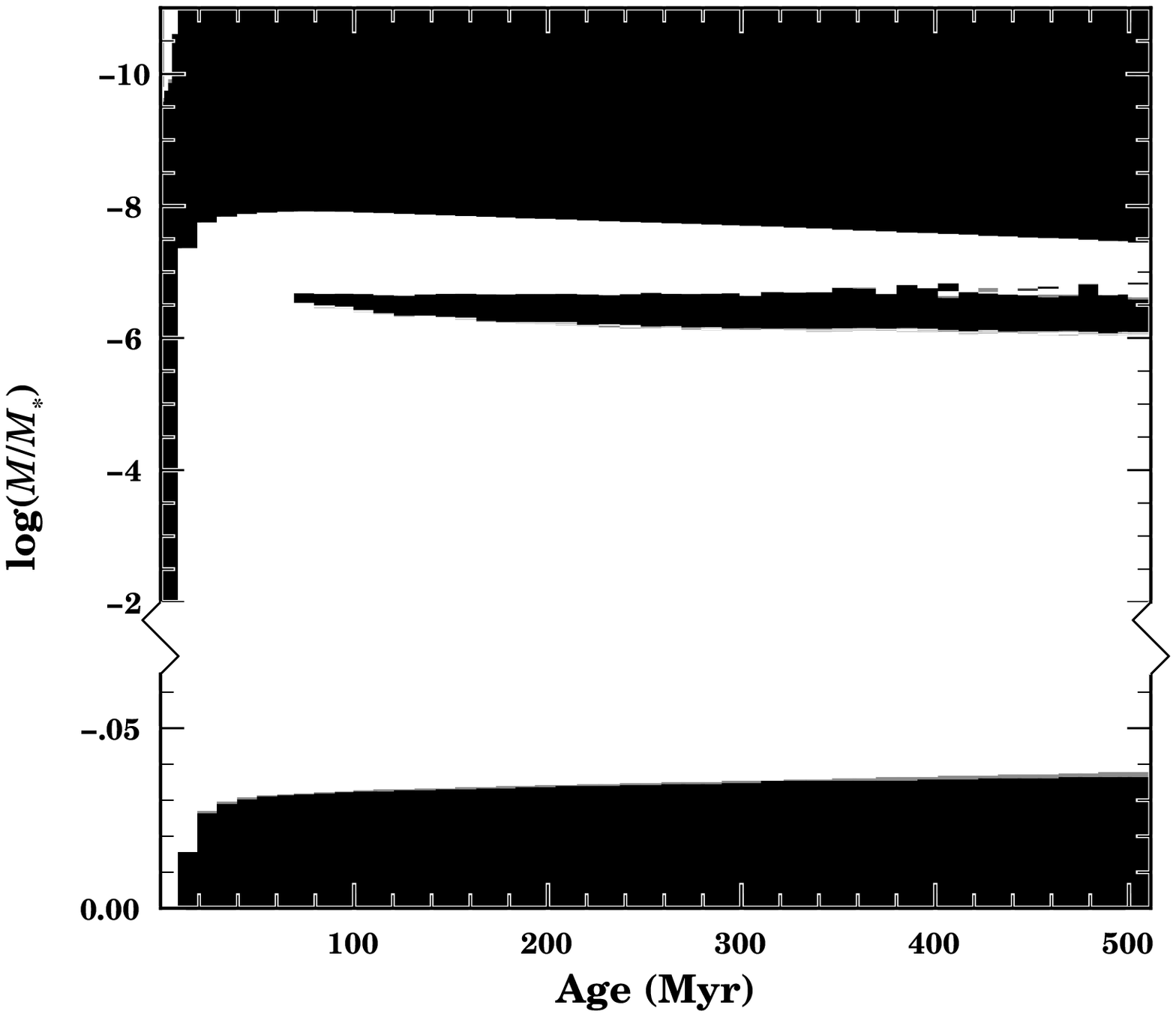}
\includegraphics[scale=.30]{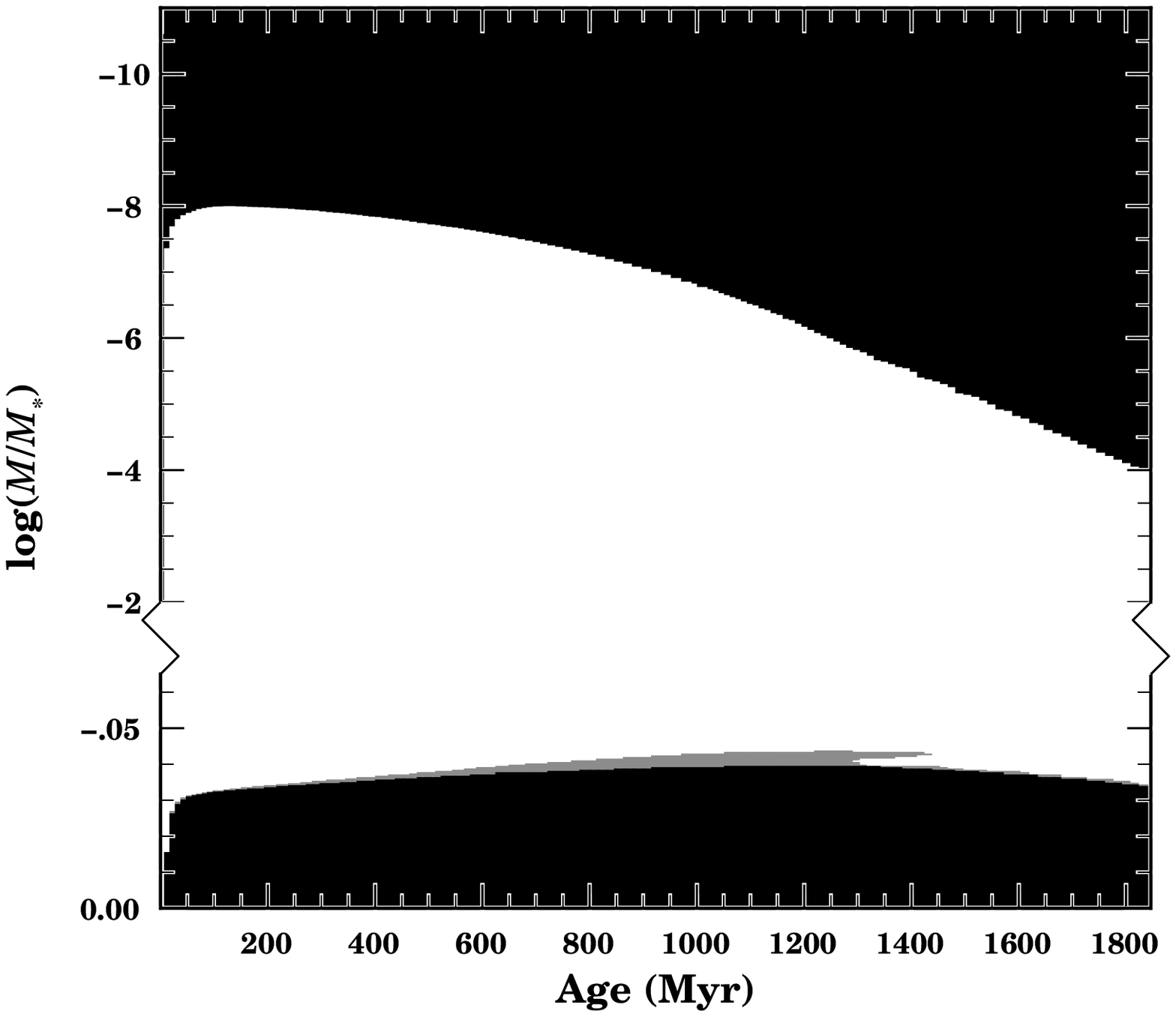}
\caption{\footnotesize
Evolution of convection (black) and semiconvection (gray) zones in two 1.5\,M$_{\odot}$ models (1\,$\times$\,10$^{-14}$M$_{\odot}$yr$^{-1}\,[right]$;\,2\,$\times$\,10$^{-14}$M$_{\odot}$yr$^{-1}$\,[left]). In the left panel, we see that an Fe convection zone develops under the main H-He convection zone. In the right panel, this convection zone does not appear because the stronger wind flattens the opacity spike due to Fe accumulation which is responsible for the Fe convection zone.}
\end{center}
\end{figure*}
The 1.5\,M$_{\odot}$ model is particularly interesting because it corresponds to the minimum mass at which iron accumulation due to the spatial distribution of its radiative acceleration causes the appearance of a convection zone (Figure 3, left panel). As mentioned above, recent evolutionary models (Richer et al. 2000, Richard et al. 2001 and Michaud et al. 2005) have successfully reproduced observations by considering that turbulence homogenizes abundances from the surface down to a temperature that corresponds to the bottom of this iron convection zone ($\log T=5.3$). Therefore, in this context, separation must take place deeper than the classical model in which separation occurs above this convection zone. Both of these scenarios have been tested with mass loss.     

In Figure 2 we have compared 6 different models of 1.5\,M$_{\odot}$ to the observed abundances of the star $\tau$UMa (Hui-Bon-Hoa 2000) 
from the Ursa Major moving group which has an age of approximately 500\,Myr (Monier 2005) and \teff $\sim$\,7000\,K (van't Veer-Menneret et M\'egessier 1996). There are 4 models in which separation is allowed immediately below the H-He convection zone (3 mass loss models, and one diffusion only model), as well as 2 models in which mixing was enforced to $\log T=5.3$ or deeper (which is the case for the turbulence model). 

It appears from the plot that the two models which best fit the data are the turbulence model as well as the model with a mass loss rate of 1\,$\times$\,10$^{-14}$M$_{\odot}$yr$^{-1}$ with enforced mixing down to the bottom of the Fe convection zone.
However the model with a mass loss rate of 2\,$\times$\,10$^{-14}$M$_{\odot}$yr$^{-1}$ and without any turbulent mixing does very nearly as well. We also see that the model with the mass loss rate 1\,$\times$\,10$^{-14}$M$_{\odot}$yr$^{-1}$ without homogenized abundances between convection zones can reproduce 
quite effectively 3 of the 5 observed abundances. It would therefore be premature to rule out the validity of the classical scenario in the context of mass loss. As we can see in the same plot, observations between Al and Ar would help in identifying if the zone between the H-He and Fe convections zones is mixed. Another important result is that turbulence and mass loss models have noticeable element to element differences which is necessary if we wish to eventually constrain the importance of both these processes. Finally, the models with the mass loss rate of 10$^{-13}$M$_{\odot}$yr$^{-1}$  flatten the abundance profiles in such a way that they can no longer reproduce observations (see also Cayrel et al. 1991). 

\section{Conclusions}
Our results seem to suggest that the scenario in which separation is to take place  at $T \sim 2\times 10^5$\,K in Fm stars must be favored over the classical scenario. In this framework, a mass loss rate of the order of the solar mass loss rate is able to reduce predicted anomalies to the observed abundances of $\tau$UMa. However, it is too early to eliminate the possibility of separation below the H convection zone. Abundance determinations between Al and Ar could help distinguish between these two regimes. It is also shown that turbulence and mass loss affect anomalies differently, though the discrepancy is slight in the models shown. Once again, more observations are required to further constrain these two mechanisms. More massive models in the \teff\,\,range where observations are not as scarce are also needed. They are currently being calculated. In any case, it is shown that reasonable mass loss rates can effectively reduce the anomalies predicted by atomic diffusion models to the observed levels.

\acknowledgements
This research was partially supported at the Universit\'e de Montr\'eal
by the Natural Sciences and Engineering Research Council of Canada. The computational resources
were provided by the R\'eseau Qu\'ebecois de Calcul de Haute Performance (RQCHP). M.Vick also thanks the 
D\'epartement de physique at l'Universit\'e de Montr\'eal as well as the GRAAl at l'Universit\'e Montpellier II.



\end{document}